\documentclass[conference]{IEEEtran}

\usepackage{amsmath,amssymb,epsfig,cite,footnote,authblk}

\newcommand{\pr}[1]{\left( #1\right)}
\newcommand{\prr}[1]{\left[ #1 \right]}
\newcommand{\es}[1]{\begin{equation}\begin{split}#1\end{split}\end{equation}}

\newcommand{\R}{\mathbb{R}}

\newcommand{\V}{\mathcal{V}}

\newcommand{\erfc}{\mathrm{erfc}}
\newcommand{\br}{\mathbf{r}}
\newcommand{\bt}{\mathbf{t}}
\newcommand{\dd}{\textrm{d}}

\begin{document}

\title{Location, location, location: Border effects in interference limited ad hoc networks}
\author[1]{Orestis Georgiou}
\author[1]{Shanshan Wang}
\author[1]{Mohammud Z. Bocus}
\author[2]{Carl P. Dettmann}
\author[3]{Justin P. Coon}
\affil[1]{Toshiba Telecommunications Research Laboratory, 32 Queens Square, Bristol, BS1 4ND, UK}
\affil[2]{School of Mathematics, University of Bristol, University Walk, Bristol, BS8 1TW, UK}
\affil[3]{Department of Engineering Science, University of Oxford, Parks Road, OX1 3PJ, Oxford, UK}
\maketitle


\begin{abstract}
Wireless networks are fundamentally limited by the intensity of the received signals and by their inherent interference.
It is shown here that in finite ad hoc networks where node placement is modelled according to a Poisson point process and no carrier sensing is employed for medium access, the SINR received by nodes located at the border of the network deployment/operation region is on average greater than the rest.
This is primarily due to the uneven interference landscape of such networks which is particularly kind to border nodes giving rise to all sorts of performance inhomogeneities and access unfairness.
Using tools from stochastic geometry we quantify these spatial variations and provide closed form communication-theoretic results showing why the receiver's location is so important.
\end{abstract}

\section{Introduction \label{sec:intro}}

With the proliferation of the internet of things (IoT), more and more devices nowadays have transceiving capabilities connecting them to the cloud via some heterogeneous wireless network backhaul.
Wireless sensor networks (WSN) is one such paradigm composed of a set of spatially distributed wireless sensor nodes tasked with monitoring some physical properties such as temperature, pressure, humidity, etc.
Data collected is then wirelessly routed through the network in a multihop fashion for storing, processing, and control purposes.
WSNs have naturally found use in environmental, military, civilian and industrial applications and are one of the main enablers of the smart city future vision.
The common Hertzian medium shared by many wireless devices however can easily become overly crowded resulting in co-channel interference and severe packet losses which need to be catered for with suitable retransmission mechanisms wasting away valuable battery juice while also incurring end-to-end delays and additional signalling overheads.

Due to the broadcast nature of wireless transmissions, concurrent signals will mutually interfere at the receiver end.
The SINR (signal to interference plus noise ratio) model has thus been widely adopted to aid in the analysis and design of efficient medium access control (MAC) protocols and communication systems \cite{haenggi2009stochastic,bachir2010mac}.
Under the SINR model, a pair of nodes can successfully communicate depending on the strength of the received (desired) signal, the background noise, and the interference caused by all other (unwanted) concurrent transmissions as measured at the receiver in question.

\begin{figure}[t]
\centering
\includegraphics[scale=0.3]{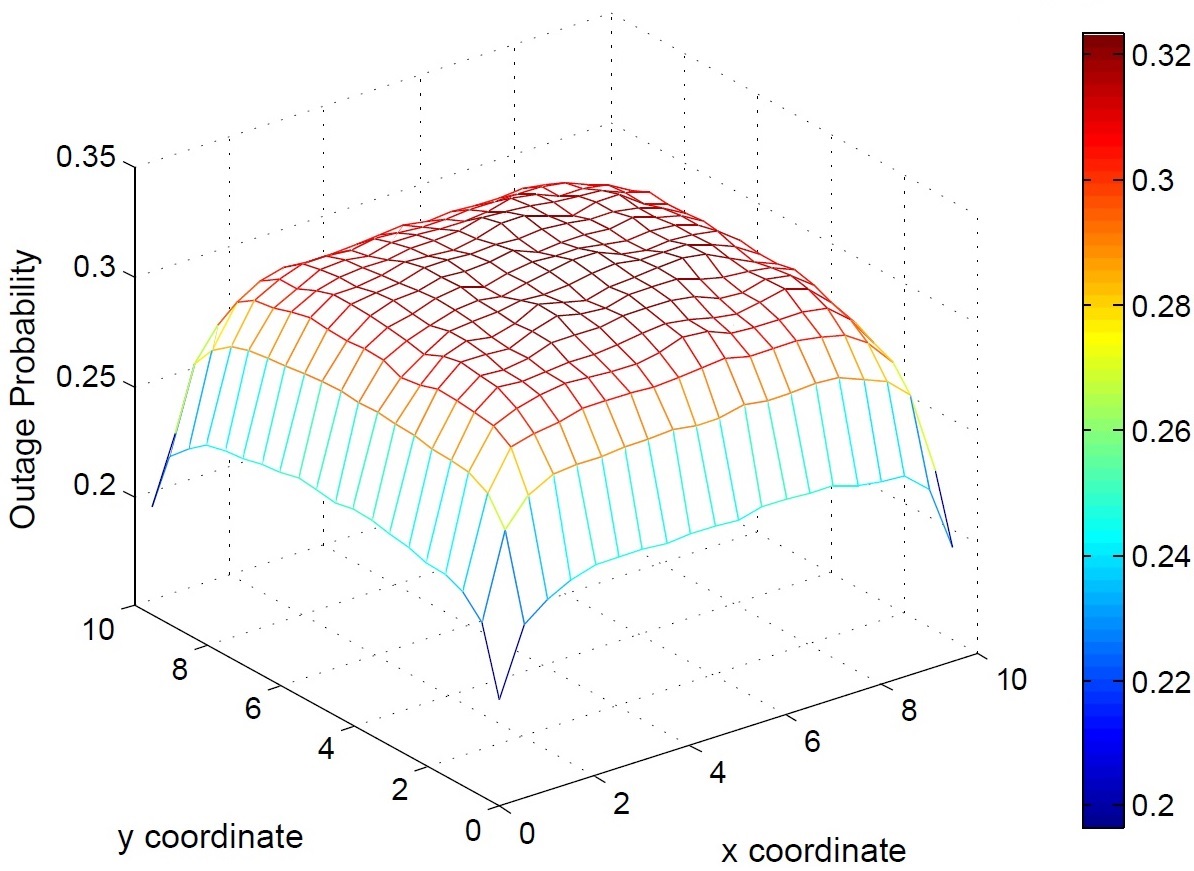}
\caption{
Plot of the outage probability for different locations of a receiving node within a $10 \times 10$ square network deployment region. In all instances the desired transmitter is at a constant distance from the receiver, and the $N_T -1$ other concurrent transmitters are uniformly distributed within the square domain.}
\label{fig:outage}
\end{figure}

Implicit in the SINR model are the numerous sources of randomness which  in some instances can be treated under a statistical framework enabling mathematical tractability and thus the analysis of local (e.g point-to-point link quality) and global (e.g. network capacity and coverage) performance metrics.
The first source of randomness comes from multi-path fading at the receiver due to the constructive/destructive self-interference of multiple phase-shifted copies of the transmitted signal.
These small-scale effects cause fast time variations in the channel gain which is typically modelled as an exponential random variable. 
Second comes the slow fading of the wireless channel due to shadowing.
These large-scale effects are often modelled using log-normal distributed random variables \cite{bettstetter2005connectivity}.
A third source of randomness is the number and location of the wireless devices e.g. in ad hoc or mobile networks. 
Even for wireless local area networks (WLANs) based on the IEEE 802.11 standards access points (APs) cannot be deployed in an optimal way (e.g. on a regular grid) due to various physical constraints and costs \cite{nguyen2007stochastic}. 
Therefore, even for planned networks, some randomness in AP locations is inherently present.
A well accepted model for the spatial node distribution is that obtained through a Poisson point process (PPP) with intensities depending on the mobility model adopted \cite{gong2014interference}.
The fourth one is power control, which helps in the management of interference but may require some cooperation in the form of signalling overheads.
Last but not least comes the channel access scheme adopted. ALOHA \cite{metcalfe1976ethernet} and CSMA \cite{kleinrock1975packet} are two well accepted classes of random and distributed MAC protocols.
The former is modelled by assuming that each node transmits randomly, irrespective of any nearby transmitter thus independently thinning the spatial PPP node distribution.
The more advanced CSMA and CSMA/CD (carrier sense multiple access with collision detection) contention protocols impose a minimum separation distance among concurrent transmitters \cite{yang2012connectivity}. 
A Mat\'{e}rn hard-core point process \cite{haenggi2009stochastic} is usually used to model the set of concurrent transmitters in such a case and in some instances can be safely approximated by the appropriately thinned parent PPP \cite{haenggi2011mean}.

While the CSMA protocol is fairly easy to understand at a local level, the
interaction among interfering nodes gives rise to quite intricate behaviour and
complex throughput characteristics on a macroscopic scale \cite{van2010insensitivity}.
Indeed, topological inequalities in the network are the cause of the observed channel access unfairness in IEEE 802.11 where nodes at the border are typically favoured.
These border effects also affect, to a certain extent, the nodes inside the network, depending on various parameters of the protocol and on the dimension of the network \cite{durvy2008border}.
This can be clearly seen in Fig. \ref{fig:outage} showing the variation of the outage probability of a receiver placed at different locations within a $10 \times 10$ network deployment region with the receiver being less likely to be in outage if placed at one of the corners of the square domain rather than in the bulk.
Similar topological variations are observed in battery life and network traffic of WSNs \cite{bachir2010mac} as well as cellular systems \cite{andrews2011tractable,novlan2013analytical} suggesting that border effects are systemic and that routing, MAC, and retransmissions schemes need to be smart, i.e. location and interference aware.

Simplifying the SINR model by ignoring fading and interference effects, one is left with the so called unit disk model (UDM) where nodes connect iff their Euclidean distance is less than or equal to some threshold, also known as the transmission range. 
Under the UDM, Penrose \cite{penrose2003random}, Gupta and Kumar \cite{gupta1998critical} proved a number of results regarding the necessary transmission range for the network to be fully connected (i.e. a multihop path exists between any two nodes) with probability approaching one as the number of nodes goes to infinity.
Mao \textit{et al.} have generalized these results for non-increasing connection probabilities \cite{mao2012towards} thus encompassing the plethora of fading and shadowing channels.
Dousse \textit{et al.} \cite{dousse2005impact} took a percolation approach to the SINR model to analyse the impact of interference and the existence (or not) of a giant connected component.
Haenggi \textit{et al.} \cite{haenggi2009stochastic} employ tools from stochastic geometry \cite{chiu2013stochastic} to treat interference as a shot noise process causing an explosion of new results and performance metrics in the realms of WSNs and cellular systems \cite{baccelli2006aloha,andrews2010random,novlan2013analytical,schilcher2014interference}.

Understanding border effects in networks has proven a difficult task. 
The main reason for this is that infinite systems are much easier to treat mathematically than finite ones.
Indeed, to the best of the authors knowledge, the move from infinite to finite networks was first achieved by Coon \textit{et al.} \cite{coon2012impact,coon2012full} and Dettmann and Georgiou \cite{dettmann2014connectivity} with rigorous results from Mao and Anderson \cite{mao2012connectivity}, alas ignoring any interference effects.

In this paper we apply tools from stochastic geometry to analyse local performance metrics under the SINR model in \textit{finite} Poisson network deployments, i.e. the nodes are placed according to a PPP in a finite subset of $\R^2$, and no carrier sensing is employed for medium access.
The main contributions of this paper are:
\begin{enumerate}
\item{We show that the interference experienced by a receiver in the network is strongly dependent on its relative location within a finite network deployment region.
In fact, we show that the location of the receiver is of equal importance as the total number of concurrent transmitting devices.
This observation is in stark contrast with previous results which consider infinite networks on the plane where all receivers are treated as being the same.
} 
\item{We quantify the above claims and provide closed form expressions for the link outage probability, the achievable ergodic rate, and the spatial density of successful transmissions which explicitly depend on the location of the receiver and the network deployment region shape.}
\item{We numerically validate our results through Monte Carlo computer simulations.
}
\end{enumerate}

The remainder of the paper is structured as follows: Sec. \ref{sec:model} introduces the system set-up and all relevant parameters, assumptions and our main metric of interest.
Sec. \ref{sec:outage} investigates the link outage probability between a specific pair of transmitting and receiving nodes for which closed form expressions are derived using tools from stochastic geometry.
Sec. \ref{sec:rate} and Sec. \ref{sec:spatial} utilize the expressions obtained for the outage probability to calculate respectively the expected achievable ergodic rate and the spatial density of successful transmissions at the receiver.
Sec. \ref{sec:num} provides numerical validation of the derived results via computer simulations and Sec. \ref{sec:con} concludes the paper and discusses some ideas and challenges for future research.

\section{Network Definitions and System Model \label{sec:model}}

Consider a finite two dimensional region $\V \subset \R^2$ of area $V=|\V|$ containing wireless devices (nodes).  
These are distributed according to two independent Poisson point processes (PPP), with intensities $\rho_T$ and $\rho_R$ within $\V$ for transmitters and receivers, respectively. 
Both intensities are zero outside $\V$.  
The number of transmitters (receivers) $N_T$ $(N_R)$ is thus a Poisson distributed random variable with mean $\rho_T V$ ($\rho_R V$). 
Such a configuration is commonly found in WSN applications where sensors or smart meters form a random mesh topology.
We denote the locations of the transmitting and receiving nodes by $\bt_i \in \V$ and $\br_j \in \V$ for $i=1,\ldots,N_T$ and for $j=1,\ldots,N_R$ respectively such that the distance between a transmitter and the receiver is given by $d_{ij}=| \bt_i - \br_j |$.
Note that the subscript order of $d_{ij}$ is important as it denotes the distance between transmitter and receiver.

It is known that fundamental results on the connectivity and capacity of dense ad hoc networks strongly depend on the behaviour of the attenuation function \cite{dousse2004connectivity}.
Here, we model the attenuation in the wireless channel as the product of a large-scale path-loss component and a small-scale fading component. 
The former follows from the Friis transmission formula where the long time average signal-to-noise ratio at the receiver (in the absence of interference) decays with distance like $\textrm{SNR}_{ij}\propto d_{ij}^{-\eta}$, where $\eta$ is the path loss exponent usually taken to be $\eta=2$ in free space and $\eta>2$ in cluttered urban environments. We therefore define a path-loss function $g(d_{ij})$ given by \cite{gong2014interference}
\es{
g(d_{ij})= \frac{1}{\epsilon + d_{ij}^\eta} , \qquad \epsilon\geq0
\label{path}
.}
Note that $g(d_{ij})$ is unit-less as is $d_{ij}$ which is scaled by the signal wavelength or some other unit of distance.
The $\epsilon$ buffer is usually included to make the path-loss model non-singular at zero.
Finally, for the sake of simplicity and mathematical tractability, the small-scale fading component is assumed Rayleigh such that the channel gain $|h_{ij}|^2$ between transmitter $i$ and receiver $j$ is modelled by an exponential random variable of mean one.
The effects of lognormal shadowing are not included in our model, suffice to say that such large-scale fluctuations of the signal are not expected to significantly affect our qualitative analysis \cite{andrews2011tractable} which is focused on border effects at the receiver at $\br_j$.

We now turn to our main metric of interest, the signal-to-interference-plus noise-ratio ($\textrm{SINR}$) between a transmitter-receiver pair; a proxy to the average link throughput. Assuming that there is no power control in the network and all devices transmit with equal powers $\mathcal{P}$ we may write
\es{
\textrm{SINR}_{ij} = \frac{\mathcal{P} |h_{ij}|^2 g(d_{ij})}{\mathcal{N} + \gamma \mathcal{I}_j}
\label{SINR}
,}
where $\mathcal{N}$ is the average background noise power, and 
\es{
\mathcal{I}_j =\sum_{k\not=i} \mathcal{P} |h_{kj}|^2 g(d_{kj})
}
is the interference received at node $j$. 
The factor $\gamma\in[0,1]$ serves as a weight for the interference term and models the gain of the spread spectrum scheme used (if any). 
For instance, in a broadband CDMA scheme $\gamma$ depends on the code orthogonality with $\gamma=1$ corresponding to a narrow band system \cite{yang2012connectivity}.
Alternatively, when $\gamma=0$ transmissions from all $k\not=i$ devices are completely orthogonal to that from device $i$ such that there is no interference received by $j$ and $\textrm{SINR}_{ij} = \textrm{SNR}_{ij}$. 
Similarly, when $\mathcal{N}=0$ there is no noise and $\textrm{SINR}_{ij} = \textrm{SIR}_{ij}$.  

In this paper we will be concerned with the full expression of \eqref{SINR} where both $\gamma>0$ and $\mathcal{N}>0$. 
Notice however that for $\gamma\approx 1$ and $\rho_T V \gg 1$, the network's performance is mostly interference limited  due to the large number of devices contributing to $\mathcal{I}_j \gg \mathcal{N}$; a \textit{shot noise process} \cite{haenggi2009interference}. 
We will next show that $\textrm{SINR}_{ij}$ and therefore the outage probability (or link quality) between nodes $i$ and $j$, strongly depends on the relative location of the receiving node $\br_j \in \V$.

\section{Outage Analysis \label{sec:outage}}

The outage probability $P_{out}$ is a fundamental performance metric of wireless networks and has been extensively studied in both cellular and mesh topologies.
For a given pair $(i,j)$ the connection probability $H_{ij}=1-P_{out}$ is given by 
\es{
H_{ij}=\mathbb{P}[\mathrm{SINR}_{ij} \geq q]
= \mathbb{P}\Big[ |h_{ij}|^2 \geq \frac{q (\mathcal{N}+ \gamma \mathcal{I}_j)}{\mathcal{P}g(d_{ij})} \Big]
\label{H}}
and can be thought of as the probability that at any given instance of time, the link between $i$ and $j$ can achieve a target $\mathrm{SINR}$ of $q$. 
Alternatively, equation \eqref{H} is the fraction of successfully received transmissions from node $i$ to node $j$, averaged over a long period of time.
It will be understood that \eqref{H} is specific for the pair $(i,j)$ as it strongly depends on the position of the receiving node $\br_j \in \V$. 
This is a direct consequence of $\V$ being finite, and is not the case in most related studies which consider homogeneous Poisson point processes (PPPs) on $\R^2$.

Conditioning on the interference realization $\mathcal{I}_j$ and using the fact that $|h_{ij}|^2 \sim \exp(1)$, the connection probability can be expressed as \cite{baccelli2006aloha}
\es{
H_{ij} &= \mathbb{E}_{\mathcal{I}_j} \Big[ \mathbb{P} \prr{|h_{ij}|^2 \geq \frac{q (\mathcal{N}+ \gamma \mathcal{I}_j)}{\mathcal{P}g(d_{ij})} \, \Big| \, \mathcal{I}_j \Big] } \\
&= \mathbb{E}_{\mathcal{I}_j} \prr{ \exp\pr{ -\frac{q (\mathcal{N}+ \gamma \mathcal{I}_j)}{\mathcal{P} g(d_{ij})} } }\\
&= e^{-\frac{q \mathcal{N}}{\mathcal{P} g(d_{ij})}} \mathcal{L}_{\mathcal{I}_j} \pr{ \frac{q \gamma}{\mathcal{P}g(d_{ij})}}
\label{laplace}
,}
where 
\es{
\mathcal{L}_{\mathcal{I}_j}(s) &= \mathbb{E}_{\mathcal{I}_j}\prr{e^{-s\mathcal{I}_j / \mathcal{P}} } = \mathbb{E}_{|h_{kj}|^2,d_{kj}}\prr{e^{-s \sum_{k\not=i} |h_{kj}|^2 g(d_{kj}) } }  \\ 
&= \mathbb{E}_{|h_{kj}|^2,d_{kj}}\prr{ \prod_{k\not=i}^{N_T-1} e^{-s |h_{kj}|^2 g(d_{kj})}  } \\
&= \mathbb{E}_{d_{kj}}\prr{ \prod_{k\not=i}^{N_T-1} \mathbb{E}_{|h_{kj}|^2} \Big[ e^{-s |h_{kj}|^2 g(d_{kj})} \Big]   }
\label{PGF}
,}
is the Laplace transform of the random variable $\mathcal{I}_j$ evaluated at $s = \frac{q \gamma}{ g(d_{ij})}$ conditioned on the locations of nodes $\bt_i$ and $\br_j$.
In the last line of \eqref{PGF} we have used the fact that the channel gains $|h_{kj}|^2$ are independent random variables. 

The probability generating function for a general inhomogeneous Poisson point process $\Xi$ in $\R^2$ with intensity function $\lambda (\xi)$ \cite{streit2010poisson} satisfies 
\es{
\mathbb{E} \Big[ \prod_{\xi \in \Xi} f(\xi) \Big] = \exp\pr{ -\int_{\R^2} (1-f(\xi) ) \lambda(\xi) \dd \xi  }
\label{PGF2},}
for functions $f$ such that $0<f(\xi)\leq 1$.
Using \eqref{PGF2} and defining $\lambda(\xi) = \rho_T$ for $\xi\in\V$ and zero otherwise, we may calculate the Laplace transform \eqref{PGF} to arrive at
\es{
\mathcal{L}_{\mathcal{I}_j} ( s ) = \exp\pr{ - \rho_T \!\! \int_\V \!\Big(1- \mathbb{E}_{|h_{kj}|^2} \Big[ e^{-  s  |h_{kj}|^2 g(d_{kj}) } \Big]\Big) \dd \bt_k } 
\label{sigma}
}
where we have assumed that the channel gains $|h_{kj}|^2$ from all simultaneous transmissions received at $\br_j$ are identically and independently exponentially distributed such that
\es{
f(\bt_k)=\mathbb{E}_{|h_{kj}|^2} \Big[ e^{- \zeta |h_{kj}|^2 } \Big]\! =\! \int_{0}^{\infty} \!\! e^{- z(\zeta +1)} \dd z = \! \frac{1}{1+ \zeta}
,}
where $\zeta= s  g(d_{kj})$ and $s = \frac{q \gamma}{ g(d_{ij})}$.
We therefore have that
\es{
\mathcal{L}_{\mathcal{I}_j} ( s ) = \exp\pr{-\rho_T\int_\V \frac{ s  g(d_{kj})}{1+ s  g(d_{kj})} \dd \bt_k }
\label{sigma2}
.}
Note that the integral in \eqref{sigma2} is over a finite domain and is well behaved.
In order to evaluate it however we must first define $\V$ and also specify the exact location of $\br_j$.

\textit{\textbf{Remark 1:} Equation \eqref{sigma} is the main result of this paper.
It describes the probability that two stations located at $\bt_i$ and $\br_j$  can achieve a target SIR of $q$ in the presence of approximately $N_T-1\gg 1$ interfering shot noise signals whose sources are uniformly distributed nodes in the finite domain $\V$.
Note that the fading of the interfering signals need not be Rayleigh, $\bt_i$ and $\br_j$ need not be inside $\V$, and finally $\V$ need not be a convex shape.
}

For the sake of juxtaposition, consider a circular sector domain $\V$ of radius $R$ and arclength $R\theta$ centred at the origin, and let $\br_j = 0$.
Hence $d_{kj}=|\bt_k|=t_k$ and the integral in \eqref{sigma2} can be evaluated using polar coordinates to give
\es{
I_R(\eta)&= \int_\V \frac{ s  g(t_k)}{1+ s  g(t_k)} \dd \bt_k = \theta \int_{0}^{R}
\frac{ s  g(t_k)}{1+ s  g(t_k)} t_k \dd t_k \\
&= \frac{\theta R^2  s }{2(\epsilon+ s )} {}_2F_1 \pr{1,\frac{2}{\eta},\frac{2}{\eta}+1,\frac{-R^{\eta}}{\epsilon+ s }}
\label{integral}
,}
where ${}_2F_1$ is the Gauss hypergeometric function. 
Equation \eqref{integral} has simpler closed form expressions
\es{
I_{R}(2)=\frac{\theta s }{2} \ln\pr{1 + \frac{R^2}{\epsilon +  s }}
} 
\es{
I_{R}(4)=\frac{\theta  s }{2\sqrt{\epsilon +  s }} \arctan \frac{R^2}{\sqrt{\epsilon +  s }}
,} 
for $\eta=2$, and $\eta=4$ respectively. 
In the limit of $R\to\infty$ equation \eqref{integral} becomes 
\es{
I_{\infty}(\eta)=\frac{\theta\pi s  (\epsilon+ s )^{\frac{2}{\eta}-1}}{\eta\sin \frac{2\pi}{\eta}}
\label{inf}
,}
where $s = \frac{q \gamma}{g(d_{ij})}$.
The infinite network assumption requires $\eta>2$ to avoid an \textit{interference storm} where $I_\infty(2) \to \infty$ and $H_{ij}\to 0$, and is reminiscent of Olbers' paradox (dating all they way back to Kepler in 1610) also called the ``dark night sky paradox'' stating that the darkness of the night sky conflicts with the assumption of an infinite and eternal static universe.
While there are a number of explanations to this very old paradox, it is clear that wireless networks are not infinite and so equation \eqref{inf} is an upper bound to \eqref{integral}, indeed  a good approximation when $R$ and $\eta$ are large.
The transition at $\eta=2$ is between local and global behaviour of the network, and is where interference starts depending more on the overall geometry than the local features.
Hence we expect that in the 3D generalisation of this work, the transition would occur at $\eta=3$.

Putting everything together we arrive at
\es{
H_{ij}= e^{-\frac{q \mathcal{N}}{\mathcal{P} g(t_i)}}e^{-\rho_T I_R (\eta)}
,
\label{H2}}
where we used $d_{ij}=|\bt_i|=t_i$.
Note that $I_{R}(\eta)$ has units of area as can also be seen from the closed form expression of \eqref{integral} making the exponentials in \eqref{H2} unitless and thus $H_{ij}\in[0,1]$ a probability. 
It follows that the connection probability depends exponentially on $\theta$, the angle over which interfering signals can arrive at $\br_j$.
Namely, receiving nodes with small $\theta$ (i.e. near the border of the domain $\V$) will typically receive less interference, and thus are less likely to be in outage than receiving nodes located in the bulk of the domain.
This shows that the location of the receiver $\br_j$ is of equal importance as the total number of transmitting devices $N_T$.
The above description therefore agrees with the spatial variation of the outage probability obtained through numerical simulations depicted by Fig. \ref{fig:outage}.
Moreover, the quantified border effects captured via $\theta$ may be regarded as a form of spatial filtering which is agnostic to the specifics of both ALOHA and CSMA-like protocols \cite{fabbri2010impact}.
We will further confirm the accuracy of our analytical results through extensive numerical simulations in Sec. \ref{sec:num}.

If the exact number of transmitters is known, then we are left with a Binomial point process (BPP) such that
\es{
&\mathbb{E}_{d_{kj}} \prr{ \prod_{k\not=i}^{N_T-1} f(\bt_k) \, \Big| \, N_T=N } = \prod_{k\not=i}^{N_-1} \mathbb{E}_{d_{kj}}\prr{  f(\bt_k) } \\
&= \pr{\mathbb{E}_{d_{kj}}\prr{  f(\bt_k) }}^{N-1} = \pr{\frac{1}{V}\int_\V \frac{1}{1+s g(d_{kj})} \dd \bt_k}^{N-1}
}
since node locations are independent random variables.
Therefore, the connection probability between two nodes $i$ and $j$ with exactly $N-1$ uniformly distributed interfering transmitters inside a domain $\V$ is given by
\es{
\mathbb{P}\prr{\textrm{SINR}_{ij}\geq q \big | N} = e^{-\frac{q \mathcal{N}}{\mathcal{P} g(d_{ij})}} \!\pr{\frac{1}{V}\int_\V \frac{1}{1+s g(d_{kj})} \dd \bt_k}\!^{N-1}
\label{BPP}}
which clearly also depends on the location $\br_j$ of the receiver.

We now turn to a similar performance metric based on $\mathrm{SINR}_{ij}$, namely the average achievable ergodic rate $\tau_{ij}$ supported by the pair in question.

\section{Achievable Rate Analysis \label{sec:rate}}

The mean data rate between two stations $i$ and $j$ in units of nats/Hz (1 bit = $\ln2 \approx 0.693$ nats) assuming adaptive modulation and coding is used is given by the Shannon capacity $\tau_{ij}=\mathbb{E}[\ln (1+\mathrm{SINR}_{ij})]$ where the average is taken over the spatial distribution of the interfering transmissions and the fading distribution.
For positive random variable $\mathbb{E}[X] = \int \mathbb{P}[X>x] \, \dd x$. 
It therefore follows that 
\es{
\tau_{ij} &= \int_{0}^{\infty}  \mathbb{P}\prr{\ln\pr{1 + \frac{\mathcal{P} |h_{ij}|^2 g(d_{ij})}{\mathcal{N} + \gamma \mathcal{I}_j}}>x} \dd x \\
&= \int_{0}^{\infty} e^{-\frac{\hat{q} \mathcal{N}}{\mathcal{P} g(d_{ij})}} \mathcal{L}_{\mathcal{I}_j} \pr{ \frac{\hat{q} \gamma}{\mathcal{P} g(d_{ij})}} \dd x \\
&= \int_{0}^{\infty} \! \exp\pr{-\frac{\hat{q} \mathcal{N}}{\mathcal{P} g(t_i)}} \exp\Big(-\rho_T \frac{\theta R^2  s }{2(\epsilon+ \hat{s} )} \\
&\times {}_2F_1 \pr{1,\frac{2}{\eta},\frac{2}{\eta}+1,\frac{-R^{\eta}}{\epsilon+ \hat{s} }} \Big) \dd x
,
\label{rate}}
where $\hat{q}= e^{x}-1$ and $\hat{s} = \frac{\hat{q} \gamma}{ g(d_{ij})}$, and in the last line we have assumed that $\V$ is a circular sector domain and $\br_j =0$ as in Sec. \ref{sec:outage}.
Equation \eqref{rate} cannot be integrated analytically but can easily be evaluated numerically using standard techniques.
It is clear however, that the position of the receiving node $\br_j$ characterised by its network angular visibility $\theta$ is a dominant parameter in the integrand of \eqref{rate} and hence will affect the average ergodic rate between the two stations.
Namely, we expect that receiving nodes near the borders of $\V$ will on average achieve higher data rates.
Moreover, we expect variations and fluctuations to these data rates will be larger for border nodes since higher moments of the rate are given by $\mathbb{E}[X^\alpha]=\alpha\int x^{\alpha-1} \mathbb{P}[X>x]\dd x$, and therefore
\es{
\tau_{ij}^{(\alpha)} &= \alpha \int_{0}^{\infty} x^{\alpha-1} \mathbb{P}\prr{\ln\pr{1 + \frac{\mathcal{P} |h_{ij}|^2 g(d_{ij})}{\mathcal{N} + \gamma \mathcal{I}_j}}>x} \dd x \\
&= \alpha\int_{0}^{\infty} \! \exp\pr{-\frac{\hat{q} \mathcal{N}}{\mathcal{P} g(t_i)}} \exp\Big(-\rho_T \frac{\theta R^2  s }{2(\epsilon+ \hat{s} )} \\
&\times {}_2F_1 \pr{1,\frac{2}{\eta},\frac{2}{\eta}+1,\frac{-R^{\eta}}{\epsilon+ \hat{s} }} \Big) x^{\alpha-1} \, \dd x
\label{rate2}
.}
We will confirm the expected rate \eqref{rate} and its higher moments \eqref{rate2} with numerical simulations later on (see Fig. \ref{fig:rate}).

\section{Spatial Density of Successful Transmissions \label{sec:spatial}}

\begin{figure}[t]
\centering
\includegraphics[scale=0.19]{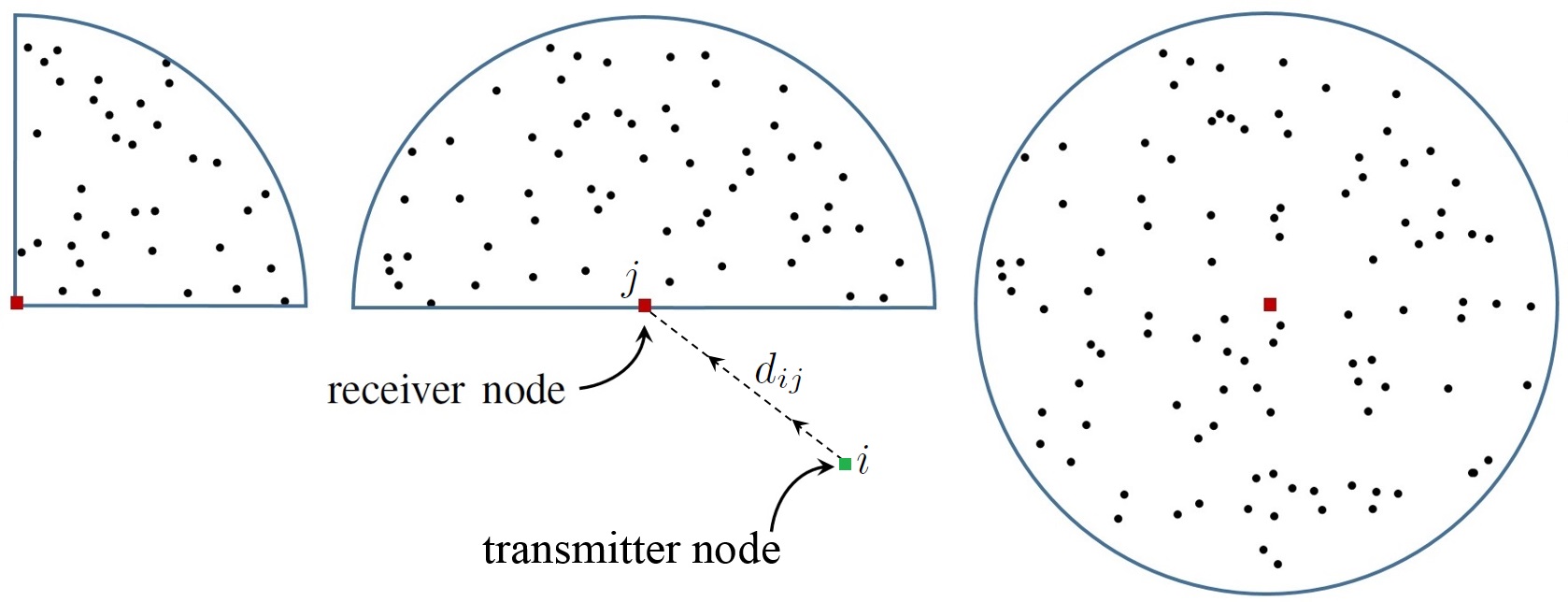}
\caption{
The three circular sector domains $\V$ of radius $R$ and angle $\theta=\pi/2,\pi$, and $2\pi$ from left to right considered in the numerical simulation that follow. The receiver node $j$ (red square) is located at the origin, the transmitting node $i$ (green square) is located a distance $d_{ij}$ away from the origin, and the $N_T-1$ interfering transmitting nodes (black dots) are uniformly distributed inside the deployment region $\V$.
}
\label{fig:example}
\end{figure}

Receiving nodes situated near the edges ($\theta\approx \pi$) and corners ($\theta<\pi$) of the domain $\V$, receive less interference and hence are (exponentially) less likely to be in outage and have higher achievable data rates.
In contrast, receiving nodes in the bulk ($\theta\approx 2\pi$) of the domain $\V$ receive more interference and hence are more likely to be in outage and have lower achievable data rates.
At the same time however, nodes near the edges and corners of $\V$ are also less likely to be the intended destination of nearby transmissions.
We must therefore examine performance metrics which are independent of the transmitting node $i$ and its location $\bt_i$.
In other words, one must average over all possible locations of $\bt_i$ and consider the spatial density of successful transmissions at $\br_j$ and not just the outage itself \cite{baccelli2006aloha}.
That is, from the $N_T$ simultaneous signals which are expected to arrive at a given receiving node $j$ located at $\br_j\in\V$, only $\mu_j$ signals are expected to achieve a target SINR of $q$ as given by
\es{
\mu_j &= \rho_T \int_\V H_{ij} \dd \bt_i \\
&= \theta \rho_T \! \!\int_0^R \!\! e^{-\frac{q \mathcal{N}}{\mathcal{P} g(t_i)}}e^{-\rho_T I_R(\eta) }  \! t_i \dd t_i
\label{mu}
,}
where in the last line we have assumed that $\V$ is a circular sector domain and $\br_j =0$ as in the previous sections.
Note that $\rho_T H_{ij}$ is a unimodal function (i.e. has a single maximum) of $\rho_T$ with a well defined maximum at  $\rho_\mathrm{max}=1/I_{R}(\eta)$ giving an interesting physical interpretation to $I_{R}(\eta)$. 
Namely, for a given receiver located at $\br_j \in \V$, then $V/I_{R}(\eta)$ gives the total number of concurrent transmissions which will maximise the number of successfully received signals at $j$ within a range of $d_{ij}$.
Increasing the density of transmitters $\rho_T$ beyond $1/I_{R}(\eta)$ will result in less successfully received signals at $j$ as more and more links will be in outage.

Assuming that $R\to\infty, \eta=4,$ and that $\epsilon=0$, we may evaluate \eqref{mu} in closed form
\es{
\mu_j= \theta \rho_T \sqrt{\frac{\pi \mathcal{P}}{16q\mathcal{N}}}\exp\pr{\frac{\gamma \mathcal{P} (\pi \theta \rho_T )^2}{64 \mathcal{N}}} \erfc\prr{ \frac{\theta\pi \rho_T \sqrt{q \gamma \mathcal{P}}}{8\sqrt{q\mathcal{N}}}}
\label{mu1}
,}
where $\erfc\prr{x}$ is the complementary error function.
Interestingly, equation \eqref{mu1} is not unimodal and in the high density limit of $\rho_T \to \infty$ converges to $\mu_j\to \frac{2}{\pi \sqrt{\gamma q}} >0$, regardless of $\theta$ or $\mathcal{N}$.
Notice that when $\gamma=0$ (i.e. no interference) $\mu_j$ diverges as expected. 
Numerically, we observe the same qualitative behaviour when using \eqref{integral} with $R<\infty$.
For any $\epsilon>0$ however this is no longer true and $\mu_j$ is unimodal in both $\rho_T$ and $\theta$; which is not surprising since $\rho_T$ and $\theta$ are of equal importance as pointed out in Sec. \ref{sec:outage}.
This last observation is particularly interesting as one may attempt to maximize $\mu_j$ by optimizing the density of transmissions $\rho_T$ in the vicinity of an intended receiver $\br_j$ based on the linear scaling relation with its angular visibility to the network $\theta$.
We will confirm these expectations (equations \eqref{mu} and \eqref{mu1}) with numerical simulations in the following section (see Fig. \ref{fig:Muij}).

\section{Numerical Simulations
\label{sec:num}}

\begin{figure}[t]
\centering
\includegraphics[scale=0.226]{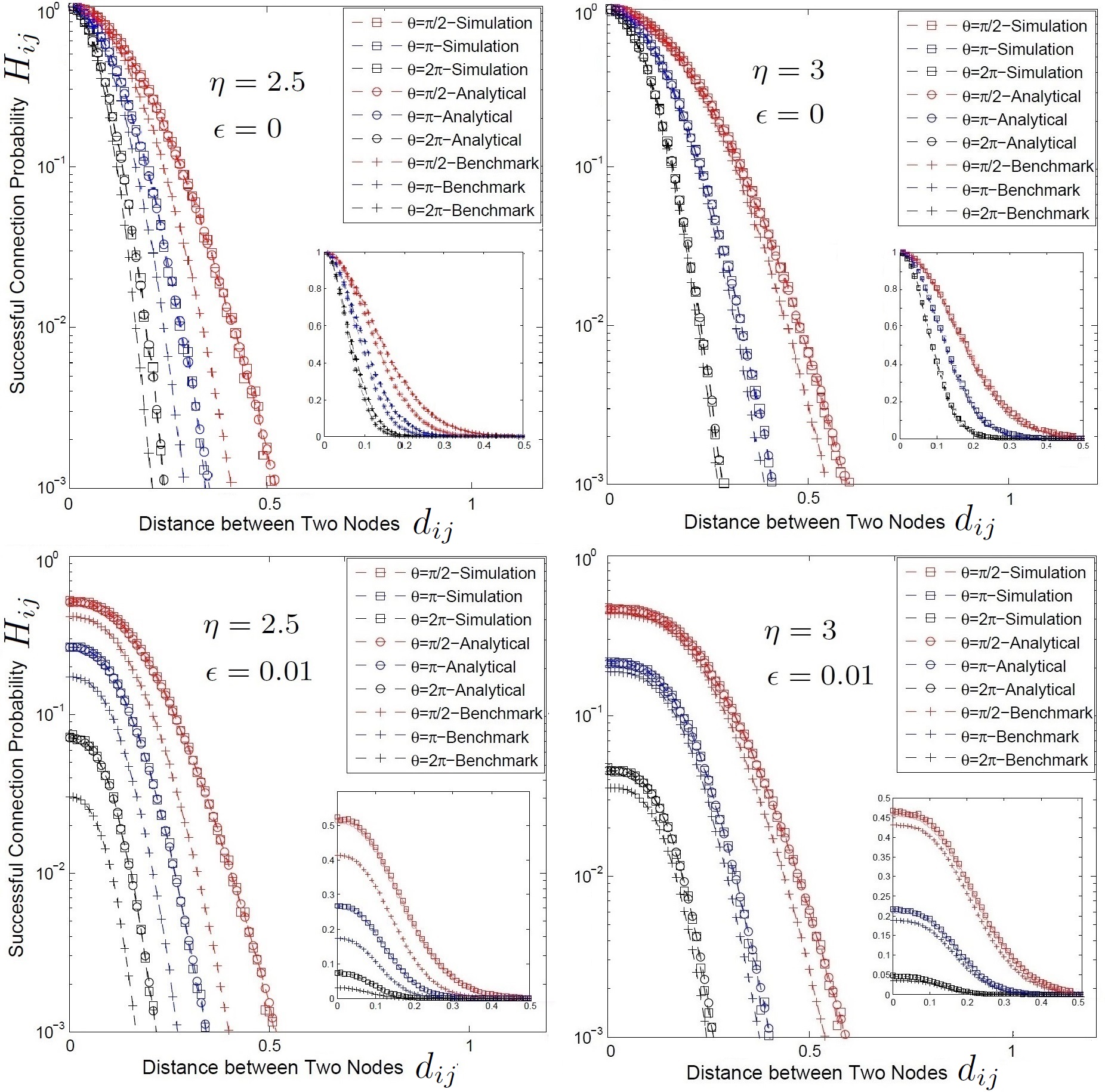}
\caption{
Log-linear plots of the connection probability $H_{ij}$ as a function of distance $d_{ij}$ between transmitter $i$ and receiver $j$ with the receiver placed at the origin of a circular sector domain of radius $R=3$ and angle $\theta=\pi/2,\pi,2\pi$, (red, blue, and black curves) using $\eta=2.5$ (left) and $\eta=3$ (right), $\epsilon=0$ (top) and $\epsilon=0.01$ (bottom).
In all cases, there are $N_T-1= \lfloor \rho_T \theta R^2 \rfloor -1$ concurrent interfering transmissions from nodes distributed randomly within $\V$ with intensity $\rho_T=12$.
An excellent agreement is observed between numerical simulations and analytical expressions \eqref{integral}.
The benchmark curve is obtained by plotting \eqref{inf}. 
The inset figures are plots on a linear axis.
}
\label{fig:conn}
\end{figure}

In the simulations that follow we uniformly and independently generate the spatial coordinates of $N_T-1\approx\lfloor \rho_T V \rfloor$ nodes in a circular sector $\V$ of radius $R=3$ and angle $\theta=\pi/2,\pi,2\pi$, using $\rho_T= 12$.
Then, we generate $N_T$ exponential random variables (modelling the channel between each transmitter and receiver $j$), and using $\mathcal{P}=\mathcal{N}=\gamma=q=1$ we calculate the $\textrm{SINR}_{ij}$ for a transmitter $i$ a distance $d_{ij}$ away from a receiver $j$ which is located at the origin of $\V$ (see Fig. \ref{fig:example} for example deployments of the three deployment regions considered).
For different distances $d_{ij}$ we repeat this process $10^5$ times in a Monte Carlo fashion using $\eta=2.5,3$ and count the number of times that $\textrm{SINR}_{ij}\geq q$ thus obtaining the connection probability $H_{ij}$.
The results are shown in Fig. \ref{fig:conn} for both $\epsilon=0$ and $\epsilon=0.01$.
We contrast the simulated results with the analytical prediction of equation \eqref{integral} and the benchmark of \eqref{inf} which assumes that the network is infinite and homogeneous.
An excellent agreement is observed between simulations and our analytical results for all three domains $\V$, while the benchmark curve is significantly off at all distances $d_{ij}$, especially when $\eta=2.5$ where it under estimates the connection probability by up to a factor of 10.
Remarkably, the lower panels in Fig. \ref{fig:conn} corresponding to $\epsilon=0.01$ have a significantly reduced connection probability $H_{ij}$ even at small distances $d_{ij}$.
This is not surprising as the strong dependence on the attenuation function $g(d_{ij})$ \eqref{path} was previously noted in \cite{dousse2004connectivity}.
Actually, there are many attenuation models which can be adopted (e.g. $g(x)=1/(\epsilon+x)^\eta$ or $g(x)=\min\pr{1,1/x^{\eta_1},1/x^{\eta_2}}$) and it is unclear which is best suited for modelling attenuation in large ad hoc networks.
It is clear however in Fig. \ref{fig:conn} and subsequent ones in this section that 
the choice of attenuation function $g(d_{ij})$ has a prominent effect especially for small station separation $d_{ij}$, which is where the near field aspects of the model come into play.
This dependence on $g(d_{ij})$ however is not the cause of discrepancy observed  between finite and infinite network regimes marked by the Analytical and Benchmark curves respectively.

\begin{figure}[t]
\centering
\includegraphics[scale=0.205]{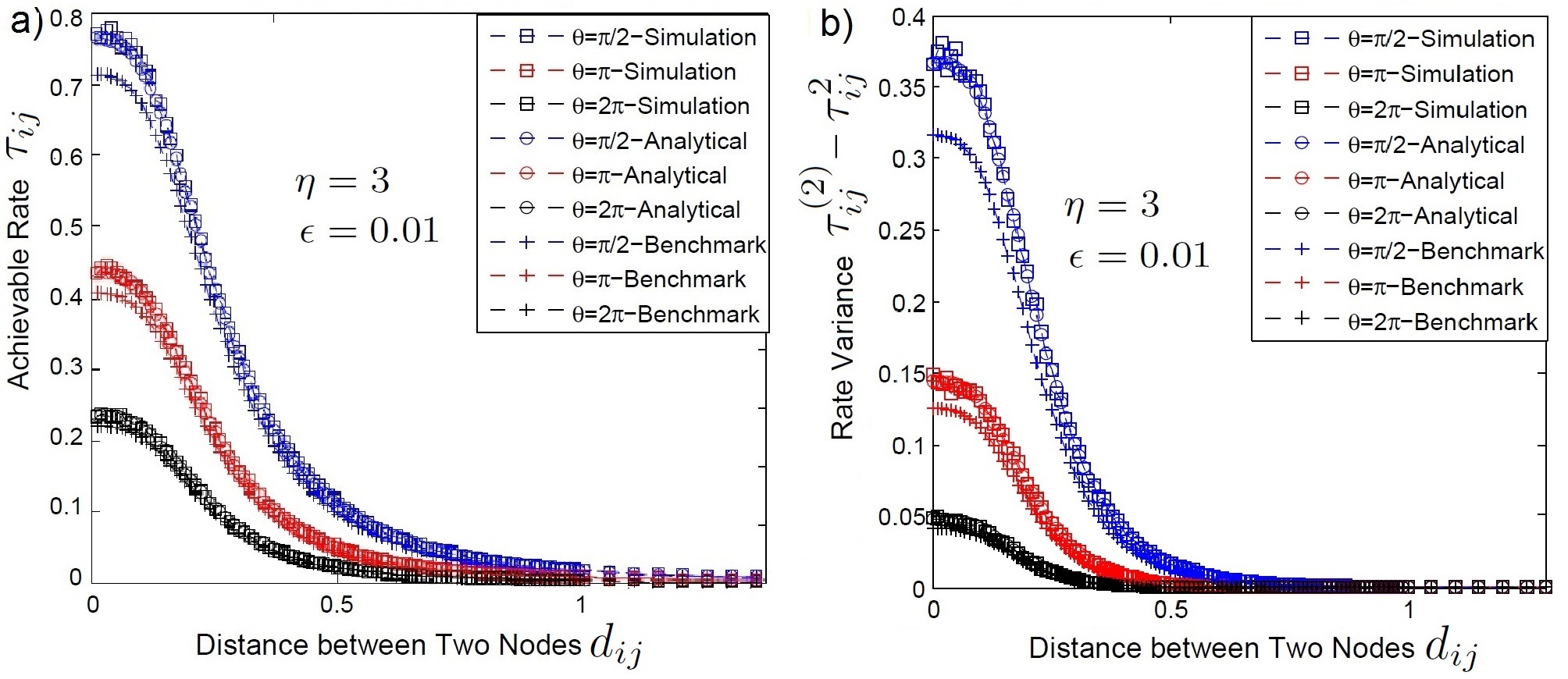}
\caption{
\textit{Left:} Plots of the mean achievable ergodic rate $\tau_{ij}$ as a function of distance $d_{ij}$ between transmitter $i$ and receiver $j$ with the receiver placed at the origin of a circular sector domain of radius $R=3$ and angle $\theta=\pi/2,\pi,2\pi,$ using $\eta=3$ and $\epsilon=0.01$. 
Meantime, there are $N_T-1= \lfloor \rho_T \theta R^2 \rfloor -1$ concurrent interfering transmissions from nodes distributed randomly within $\V$ with intensity $\rho_T=12$.
\textit{Right:} The corresponding variance of the simulated results contrasted against $\tau_{ij}^{(2)}-\tau_{ij}^2$ obtained by equation \eqref{rate2}.
}
\label{fig:rate}
\end{figure}
\begin{figure}[t]
\centering
\includegraphics[scale=0.25]{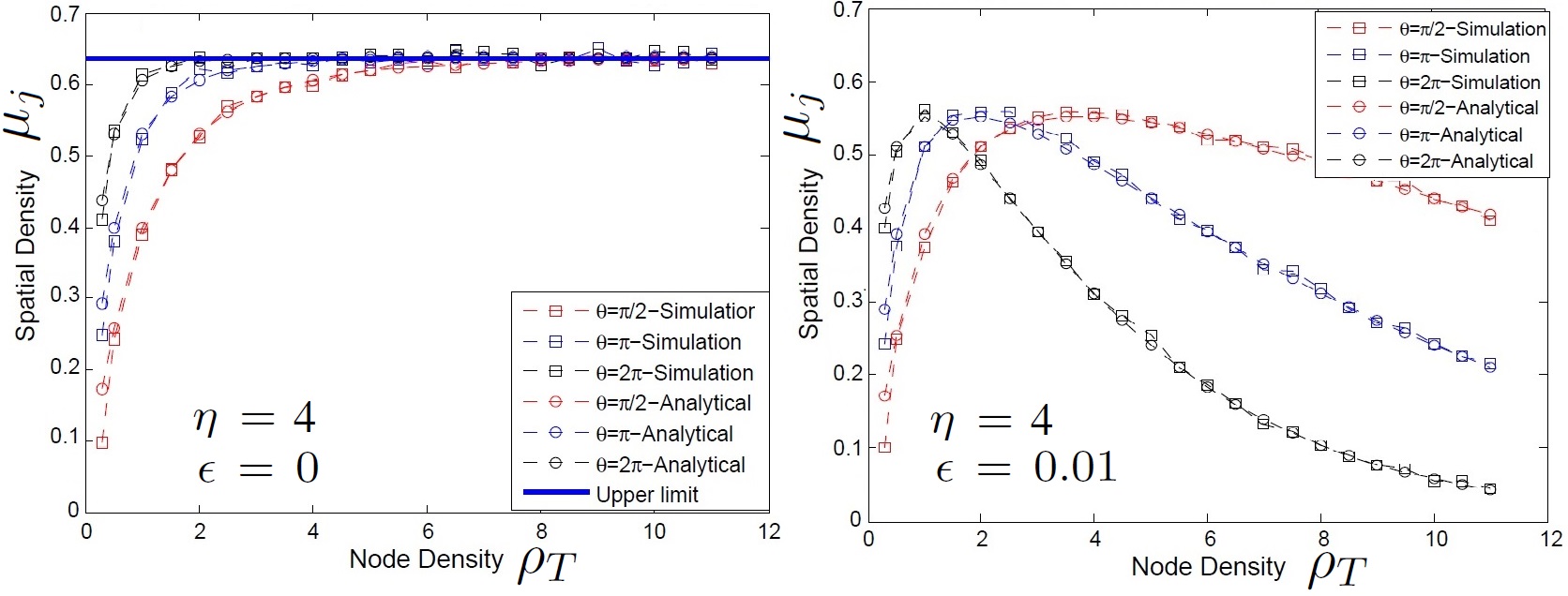}
\caption{
Plot of the spatial density of successful transmissions $\mu_j$ which can be received by a node $j$ located at the origin of a circular sector of radius $R=3$ and angle $\theta=\pi/2,\pi,2\pi,$ using $\eta=4$ and $\epsilon=0$ (left panel) and $\epsilon=0.01$ (right panel).
The analytical results are obtained through equations \eqref{mu} and \eqref{mu1}.
}
\label{fig:Muij}
\end{figure}

We next repeat the simulations and calculate the Shannon capacity $\tau_{ij}=\mathbb{E}[\ln(1+ \textrm{SINR}_{ij})]$ and its variance.
The results for the mean Shannon bound are shown in Fig. \ref{fig:rate}a) using $\epsilon=0.01$ and $\eta=3$.
We contrast the simulated results with the analytical prediction of equation \eqref{rate} and the benchmark of \eqref{inf}.
An excellent agreement is again observed illustrating the importance of considering finite networks and border effects.
The variance of the rate is plotted in Fig. \ref{fig:rate}b) which we contrast with $\tau^{(2)}_{ij}-\tau^{2}_{ij}$ obtained through the numerical integration of \eqref{rate2}.
Interestingly, the fluctuations of the rate around the mean $\tau_{ij}$ decrease with distance $d_{ij}$ and are higher for smaller values $\theta$, i.e., the rate will fluctuate more when the receiver is situated at a sharp corner of $\V$.
This is because there are fewer interfering signals (recall that $\theta$ and $\rho_T \propto N_T$ are of equal importance) and thus the total interference $\mathcal{I}_j$ at $\br_j$ will be dominated by the closest interfering node, the location of which can vary a lot with each Monte Carlo simulation run. 

Fig. \ref{fig:Muij} shows how the spatial density of successful transmissions $\mu_j$ changes with the density of interfering transmissions $\rho_T$ for different $\epsilon=0$ and $\epsilon=0.01$ using $\eta=4$.
Analytical results from \eqref{mu1} are in excellent agreement with simulations for both cases confirming the monotonicity of $\mu_j$ for $\epsilon=0$, and it's unimodality for $\epsilon>0$.
Interestingly, for $\epsilon>0$ the optimal spatial density which maximizes $\mu_j$ is higher for smaller values of $\theta$, implying that at low densities receivers located at the bulk of the domain are more likely to be the intended destination of nearby transmitters (i.e. higher $\mu_j$). 
The opposite is true at high densities as receivers located at the corners of the domain are the ones more likely to be the intended destination of nearby transmitters.

\section{Discussion and Conclusions
\label{sec:con}}

Wireless networks can suffer high interference levels due to the limited yet heavy use of spectrum leading to packet losses and end-to-end delays.
With the current densification trend (particularly in urban and indoors areas) of wireless devices and APs, it has become of paramount importance to develop mathematical models which can accurately describe interference effects and help design more efficient MAC protocols and communication systems.
The SINR model \cite{haenggi2009stochastic} is one such attempt employing tools from stochastic geometry \cite{chiu2013stochastic}, is well accepted and researched, and is currently used in the analysis of WLAN \cite{nguyen2007stochastic} and cellular \cite{andrews2011tractable,novlan2013analytical} network systems.

In this paper we have addressed the spatial variations of SINR and related performance metrics (e.g. link outage probability, the Shannon capacity, and the spatial density of successful transmissions) in interference limited ad hoc networks operating in a finite region $\V\subset\R^2$ and in the absence of carrier sensing for medium access.
In particular, we study how these metrics depend on the location of the receiver node in $\V$, and derive communication-theoretic results in closed form, thus revealing the often non-trivial relationship between system and device parameters (e.g. noise and transmit power, pathloss, bandwidth etc.) with respect to the location of the receiver.
Significantly, we show that the location of the receiver is equally important to the total number of concurrent interfering transmissions therefore suggesting that routing, MAC, and retransmissions schemes need to be smart, i.e. location and interference aware.
For example, we show that under uniform assumptions, receiver nodes near the border of $\V$ experience less interference (compared to nodes at the center of $\V$) and therefore are less likely to be in outage.
In current decentralized medium access protocols based on CSMA/CA, such as IEEE 802.11, this translates to an unfair advantage to border nodes which have a higher probability to access the communication channel \cite{durvy2008border}. 
Hence, smarter MAC and routing protocols could regard routes involving border nodes as being more reliable on average, simply due to their locations within the network mesh in an attempt to instil fairness.



\section*{Acknowledgements}

The authors would like to thank the directors of the Toshiba Telecommunications Research Laboratory for their support and the anonymous referees for their useful comments.


\bibliographystyle{ieeetr}
\bibliography{mybib}

\end{document}